\PassOptionsToPackage{square,comma,numbers,sort&compress}{natbib}
\documentclass{techmech}
\usepackage{natbib}
\usepackage{subfig}
\usepackage{graphicx}

\usepackage{xcolor}
\usepackage{mdframed}

\title{The impact of AI on engineering design procedures for dynamical systems}
\author{%
Kristin M.~de Payrebrune\textsuperscript{1$\star$$\dagger$}\!, 
Kathrin Flaßkamp\textsuperscript{2$\dagger$}\!, 
Tom Ströhla\textsuperscript{3}\!, 
Thomas Sattel\textsuperscript{3}\!, 
Dieter Bestle\textsuperscript{4}\!, 
Benedict Röder\textsuperscript{5}\!,
Peter Eberhard\textsuperscript{5}\!, 
Sebastian Peitz\textsuperscript{6}\!, 
Marcus Stoffel\textsuperscript{7}\!,
Gulakala Rutwik\textsuperscript{7}\!,
Borse Aditya\textsuperscript{7}\!, 
Meike Wohlleben\textsuperscript{8}\!, 
Walter Sextro\textsuperscript{8}\!, 
Maximilian Raff \textsuperscript{9}\!, 
C.~David Remy\textsuperscript{9}\!,
Manish Yadav\textsuperscript{10}\!, 
Merten Stender\textsuperscript{10}\!,
Jan van Delden\textsuperscript{11}\!,
Timo Lüddecke\textsuperscript{11}\!,
Sabine C.~Langer\textsuperscript{12}\!
Julius Schultz\textsuperscript{12}\! and 
Christopher Blech\textsuperscript{12}\!
}
\renewcommand{\tmauthorlist}{K.~M.~de Payrebrune et al.}
\affil{\small%
\textsuperscript{$\dagger$} \textit{Both authors contributed equally in organizing this paper. For list of contributions, see end of manuscript.} \newline
\textsuperscript{1} RPTU Kaiserslautern-Landau, Institute for Computational Physics in Engneering, Gottlieb-Daimler-Straße 74, 67663 Kaiserslautern, Germany \newline 
\textsuperscript{2} Saarland University, Systems Engineering, Systems Modeling and Simulation, Campus, 66123 Saarbrücken, Germany \newline 
\textsuperscript{3} TU Ilmenau, Mechatronics group, Max-Planck-Ring 12, 98693 Ilmenau, Germany\newline 
\textsuperscript{4} Brandenburg University of Technology Cottbus-Senftenberg, Chair of Engineering Mechanics and Vehicle Dynamics, Siemens-Halske-Ring 14, 03046 Cottbus, Germany\newline 
\textsuperscript{5} University of Stuttgart, Institute of Engineering and Computational Mechanics, Pfaffenwaldring 9, 70569 Stuttgart, Germany\newline 
\textsuperscript{6} TU Dortmund, Department of Computer Science \& Lamarr Institute, Joseph-von-Fraunhofer-Str.\ 25, 44227 Dortmund, Germany \newline
\textsuperscript{7} RWTH Aachen University, Institute of General Mechanics, Eilfschornsteinstraße 18, 52062 Aachen, Germany\newline 
\textsuperscript{8} Paderborn University, Faculty of Mechanical Engineering, Dynamics and Mechatronics, Warburger Str.\ 100, 33098 Paderborn, Germany\newline
\textsuperscript{9} University of Stuttgart, Institute for Nonlinear Mechanics, Pfaffenwaldring 9, 70569 Stuttgart, Germany\newline 
\textsuperscript{10} Technische Universität Berlin, Chair of Cyber-Physical Systems in Mechanical Engineering, Straße des 17.~Juni 135, 10623 Berlin, Germany\newline 
\textsuperscript{11} University of Göttingen, Institute of Informatics, Goldschmidtstr.~7, 37077 Göttingen, Germany \newline 
\textsuperscript{12} Technische Universität Braunschweig, Institute for Acoustics and Dynamics, Langer Kamp 19, 38106 Braunschweig, Germany \newline 
}
\renewcommand{\corremail}{kristin.payrebrune@mv.rptu.de}
\renewcommand{\tmkeywords}{Product Design Assistants, AI methods, Exemplary procedure, AI limitation}
\renewcommand{\tmabstract}{Artificial intelligence (AI) is driving transformative changes across numerous fields, revolutionizing conventional processes and creating new opportunities for innovation. The development of mechatronic systems is undergoing a similar transformation. Over the past decade, modeling, simulation, and optimization techniques have become integral to the design process, paving the way for the adoption of AI-based methods. In this paper, we examine the potential for integrating AI into the engineering design process, using the V-model from the VDI guideline 2206, considered the state-of-the-art in product design, as a foundation. We identify and classify AI methods based on their suitability for specific stages within the engineering product design workflow. Furthermore, we present a series of application examples where AI-assisted design has been successfully implemented by the authors. These examples, drawn from research projects within the DFG Priority Program \emph{SPP~2353: Daring More Intelligence - Design Assistants in Mechanics and Dynamics}, showcase a diverse range of applications across mechanics and mechatronics, including areas such as acoustics and robotics.}

\makeatletter
\newcommand\footnoteref[1]{\protected@xdef\@thefnmark{\ref{#1}}\@footnotemark}
\makeatother

\begin{document}
\pagestyle{scrheadings} 
    \maketitle


\section{Introduction} 
\label{sec:introduction}

Structural changes, shorter development cycles and increasingly interdisciplinary challenges require novel approaches to product design.
Future system design in engineering has to be \emph{multidisciplinary}, not only in order to meet future customer requirements, new regulations, and social responsibilities, but also to optimally benefit from the technical possibilities of new system types with integrated Artificial Intelligence (AI) components \cite{verganti2020innovation}.
As AI methods originate from mathematical algorithms, initially, they are not tailored to certain disciplines, but follow general principles which make them applicable to the individual disciplines as well as on integration level (cf.\ Fig.~\ref{fig:Vmodel} which sketches the V-model design process).

Future system design will only be achieved by \emph{(partially) automating the design process}: (1) numerical optimization strategies take over the search for the best compromise solutions, (2) analysis methods from different disciplines are integrated for system evaluation, and (3) AI-enhanced algorithms are applied to system dynamics evaluation and control, as well as to design decisions \cite{salehi2018emerging}.

The design of novel technological systems is one of the most demanding creative achievements in engineering.
Currently, assistance by numerical tools is limited primarily to system analysis. Hence, the typical development process is realized in an interactive nature: firstly, time-consuming parameter studies are carried out,
then, simulation results are manually and subjectively evaluated, leading to iterative changes to the current design based on expert knowledge.
The result is then often declared "optimal", which seems questionable in multiple regards: (i) in general, the evaluation is based on subjective criteria rather than on objective, mathematically described criteria; (ii) the human inability to recognize correlations in higher-dimensional design spaces as well as time constraints in the industrial development process prevent the achievement of an optimum; (iii) the expertise of a design engineer is usually limited to one or only a few sub-disciplines.

An AI-assisted design process provides the opportunity to overcome these shortcomings \cite{yousif2022towards, kwong2016ai}.
Moreover, it allows the consideration of various design criteria already in \emph{early design phases}, which used to be considered at a later stage only.
This might, at the same time, contribute to better design outcomes and to shorter development times.

\begin{figure}[bth]
    \centering
    \includegraphics[width = 0.5\textwidth]{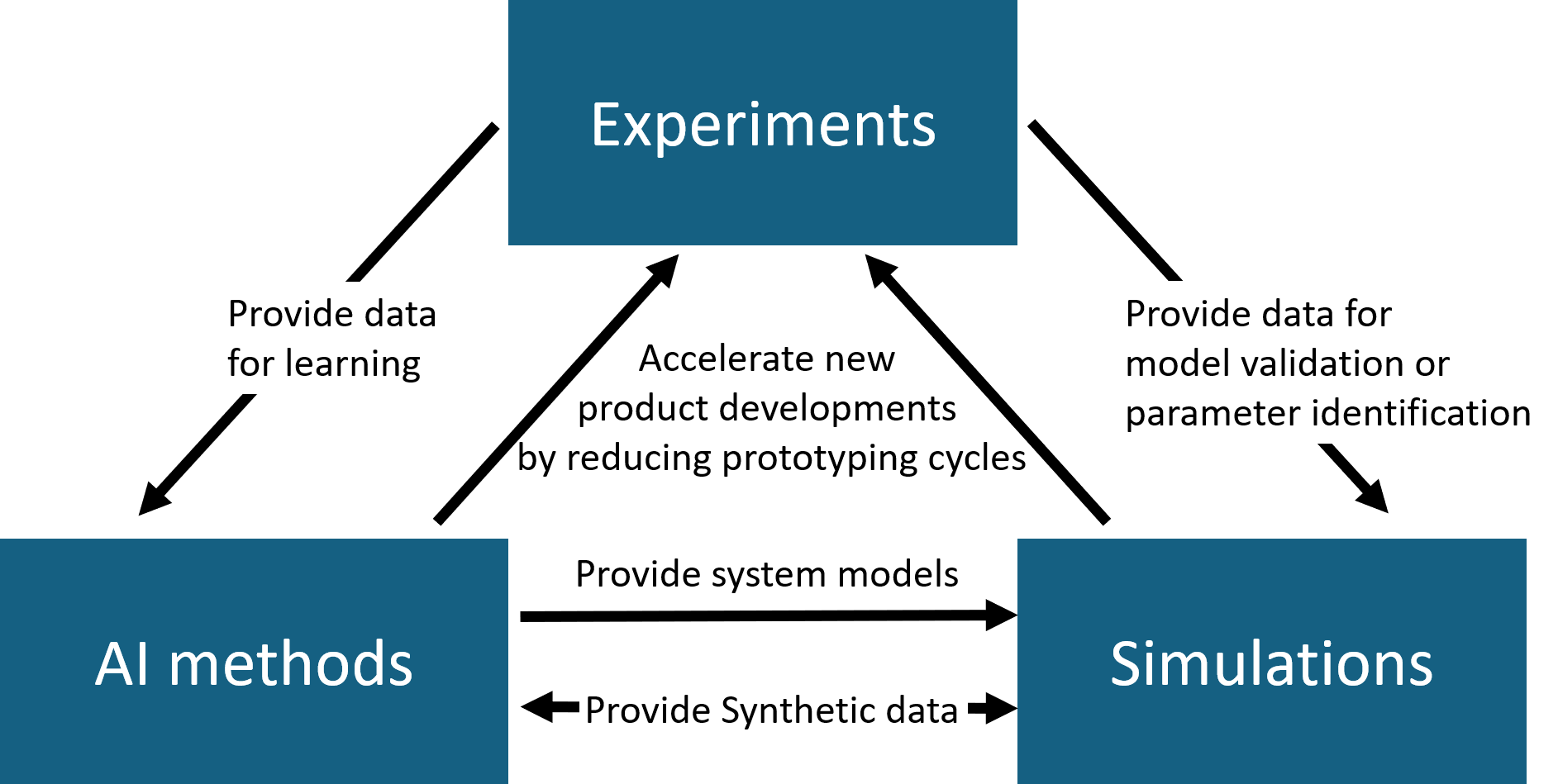}
    \caption{In mechatronic design, AI methods link closely to both experiments and simulations.}
    \label{fig:fields}
\end{figure}

Future system design methodologies thus have to orchestrate the interplay between \emph{experiments}, \emph{model-based simulations}, and \emph{AI methods} as is illustrated in Fig.~\ref{fig:fields}.
Classically, technical design was based heavily on experiments \cite{Faurre1975}.
Numerical simulation techniques for physics-based models can reduce the number of experiments, extend beyond existing prototypes, or test designs in extreme situations, which are challenging to reproduce in physical experiments. 
On the other hand, the model-based results must be validated by real experiments \cite{Ramirez2023, Vanommeslaeghe2022}. For complex systems, however, there is often a lack of suitable models, or the existing simulations are so extensive and time-consuming that systematic system analyses become impractical.
In addition, determining the required model parameters is difficult and tedious. Distorting, interfering, and obscured effects in experiments often prevent a simple interpretation \cite{eremeev2015determination, raue2011, Guse2020}. Vast amounts of measurement data request for automated and \textit{intelligent} data processing, potentially using data mining approaches.

In tandem with these approaches, the current landscape of product development is witnessing a notable influence from AI methodologies. Machine learning (ML), a branch of artificial intelligence, is employed to predict and optimize outcomes in simulated environments \cite{leha2009optimization}. These algorithms can discern complex patterns and relationships within simulation data, enabling more accurate predictions of product behavior. Additionally, AI-driven simulations facilitate automated parameter optimization, accelerating the iterative design process. The adaptability of AI enables simulations to learn from evolving data, providing a dynamic and intelligent framework for virtual testing \cite{shao2020simulation}. Moreover, in high-dimensional parametric spaces, AI methods are promising alternatives in the field of design-of-experiments.
Hence, for system design, not one single way is advantageous above the others, but a combination of available and best suited methods will enhance the design process.


\subsection{Contribution of this paper} 
While applications of AI emerge everywhere, its potential for assisting the design process of engineering products has not been fully revealed yet.
With the goal of exploring this potential, the contribution of this paper is threefold:
first, we analyze the (classical, or conventional) state-of-the-art design process to identify opportunities for integrating AI methods;
second, we explore suitable AI methods with respect to their application in engineering design and classify them by linking to the respective steps within a mechatronic product design process;
last, we provide a short overview of application examples, in which the authors successfully embed AI assistance into mechatronic design. 
These examples stem from research projects within the DFG priority program \emph{SPP~2353:  Daring More Intelligence - Design Assistants in Mechanics and Dynamics}\footnote{\label{note1}https://gepris.dfg.de/gepris/projekt/460725022}.

\subsection{Outline} 
The remainder of this paper is structured as follows. We review the classical methodology of mechatronic product design in Section~\ref{sec:currentstate}. 
In the main part of the paper, we derive potential for integrating AI into the design process in Section~\ref{sec:potentials}, in which we discuss synthetic data generation, AI-driven modeling approaches from various disciplines as well as from a holistic point of view, and AI-assisted methods in system identification, optimization, and control.
Then, we present innovative examples for integrating AI tools in mechatronic applications, see Section~\ref{sec:example}.
Concluding remarks with an outlook to open problems are given in Section~\ref{sec:conclusion}.

\section{State-of-the-art in engineering product design}
\label{sec:currentstate}

\begin{figure}[bth]
    \centering
    \includegraphics[width = 0.45\textwidth]{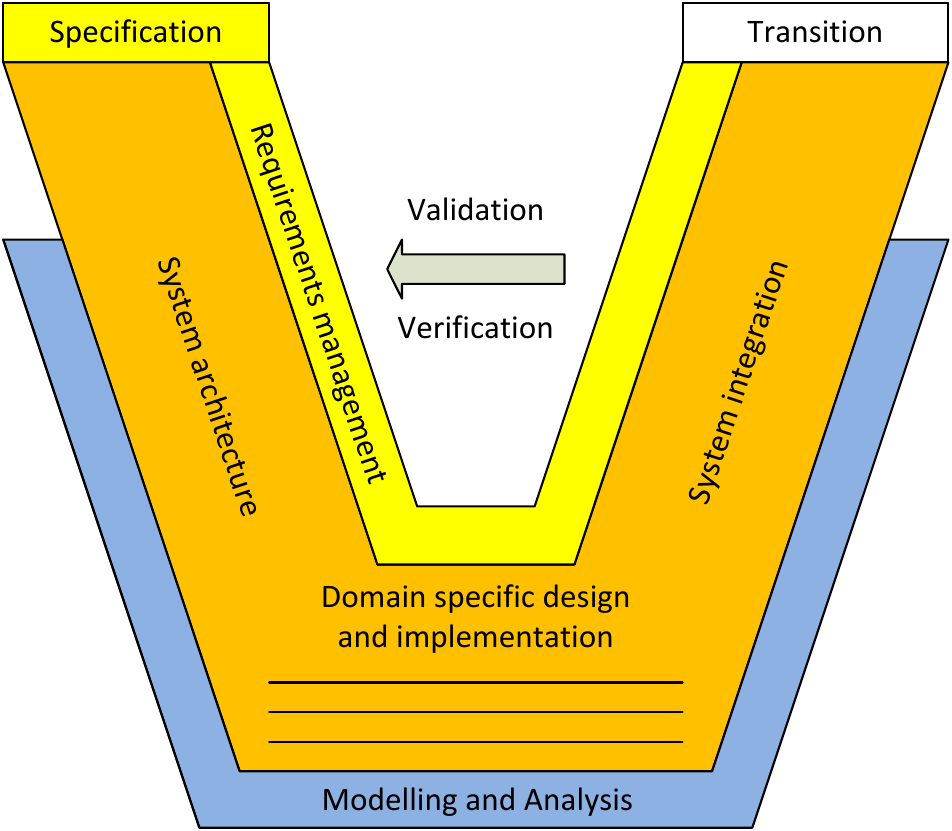}
    \caption{V-model for mechatronic design according to VDI guideline 2206 \cite{VDI2206}.}
    \label{fig:Vmodel}
\end{figure}

The VDI guidelines 2221 \cite{VDI2221} and 2206 \cite{VDI2206} were created with the aim of planning and implementing the design of technical products, in particular mechatronic systems, efficiently and reliably in terms of both time and costs. They summarize practical experience and scientific approaches. The \textit{V-model}, see Fig.~\ref{fig:Vmodel}, which originated in software development \cite{Broehl1993}, represents a widespread approach to design methodology.

The starting point of a design process is the requirements elicitation, which is documented in a \textbf{specification}. 
While, in former times, requirements stayed static during the design process, nowadays, the increased system complexity makes it impossible to fix the requirements at an early stage. Instead, flexible adaptation during the design process is implemented as feedback loops, see Fig.~\ref{fig:Vmodel}.

The left wing of the V represents the early design phases (coarse design). Here, starting with an abstract (solution-independent) representation of the task, the system concept or \textbf{system architecture} is defined by breaking it down into sub-tasks and assigning solution structures to the individual functions. Since each sub-function can be realized by many solutions from different physical domains, this results in a fast-growing solution space whose size increases exponentially with the number of sub-structures. In addition to the main functions of the solutions, secondary functions and interactions with the other sub-structures must also be taken into account. The developers' experience and creativity increase the chances of success. Usual design approaches are based on combination tables, or on simple analytical models with characteristic parameters \cite{Eigner2012}.

The \textbf{domain-specific designs} (fine design) are shown in the lower section of the V. These are carried out in parallel by specialists and have a high level of detail. In the case of mechatronic systems, a mechanical (structural) design, an electronic design, and a software design usually take place. Depending on the system design, other domains may be added, e.g., fluid dynamics, magnetics, or optics. There are special design approaches and tools for each domain. For example, computer aided design (CAD) tools or strength analyses using a finite element method (FEM) are used in mechanical design. Simulators such as a simulation program with integrated circuit emphasis (SPICE) or routing software are available for circuit design \cite{tuma2009}. Ergonomic integrated development environments (IDE), GitHub, and automatic code generation and testing tools help with software development.

The detailed domain designs are combined in the right wing of the V in the \textbf{system integration}. Thanks to the specific knowledge of all subsystems, the system behavior can be precisely predicted statically and dynamically. Furthermore, controllers can be parameterized. Analog simulations such as Matlab or Simplora help with this. The system properties are compared with the specification by means of \textbf{verification} (comparison of the system properties with the specification at different levels of concretization) and \textbf{validation} (review concerning a predefined intended use or benefit).
The design process ends with the \textbf{transition} of the design data to the production design or further design loops.

The design process is accompanied by \textbf{modeling and analysis} (simulation) on the underside of the V. As indicated in the previous description, models with different levels of detail and accuracy exist for each phase. A continuous data management is needed to process results of one step to prepare as input of the next step. This is often realized by simple scripts and databases, and hindered by proprietary software packages. The models are verified and stored with the specific parameters as a digital twin for the product, see, e.g., \cite{KW20,KPW21,TTM+24}.

The V-model has been developed by the Association of German Engineers (VDI) \cite{VDI2206}. However, similar concepts are also used in engineering cultures of other countries, e.g., in guidelines of the U.S. Department of Transportation \cite{USDepTransp2009}. It has several further developments and specific adaptations \cite{Graessler2018}, e.g., for micro-electro-mechanical systems (MEMS) \cite{Watty2006}, or for considering self-organizing production systems in early phases of interdisciplinary product development \cite{Hentze2021}.
Only recently, a revision of the V-model in the light of AI methods is proposed, see e.g., \cite{Ullrich2024}.

\section{Potential for integrating AI into the design process}
\label{sec:potentials}
AI methods already showed great success in pattern recognition and separation, text and image generation and other fields. In industrial product design, these approaches have not been used yet. However, they offer several interesting opportunities to improve the design process described in Section~\ref{sec:currentstate}.
The system architecture in Fig.~\ref{fig:Vmodel} is a synthesis process. There are hardly supporting tools for finding the best configuration of solutions for sub functions. The solution space increases exponentially with the number of sub functions. Experience and intuition of the human designers, which are mostly experts in only few technical domains can be only limited to a subspace. For instance, DMG lib,  collected more than 3000 different mechanisms partly from forgotten patents or historic panels \cite{doring2006}. Oftentimes, different domains compete at solutions for a sub function, e.g., a mechanism with standard drive vs. a combination of smart controlled drives, or a physical sensor vs. an observer in a feedback loop. AI can help by its huge database to identify suitable combinations of solutions for all sub functions and from different domains.
Also the domain specific design in Fig.~\ref{fig:Vmodel} is a synthesis process where a huge number of parameters have to be specified from only a few input parameters. On the one hand, this is an optimization process, often with a multi-criteria target function. Classical mathematical optimization approaches are limited by huge calculation times and by handling the different criteria. On the other hand, the models used in domain specific design have to be very accurate. Often they are nonlinear and contain estimated material parameters. In some cases the physical approach is valid only for a limited range of the system variables or is even completely unknown. AI approaches can support at this gap with black-box-models, which are learned from experimental and simulated data.
The integration of AI into the design process bears great potential for accelerating and transforming traditional approaches. The ability of AI to analyze huge data sets, recognize patterns and optimize solutions in real time makes it a game-changing tool for increasing efficiency and creativity in design. 

The following sections examine the current state of AI applications, and highlight the potential for improvements in the design process. Referring to the V-model in Fig.~\ref{fig:Vmodel}, we currently observed that AI methods are mostly used in the early stage development. Model-based design, which allows for a reduction of laboratory experiments and minimal prototyping, may stand in conflict with data hungry AI methods, ref. Fig.~\ref{fig:fields}. Here, synthetic data generation provide options for solutions, cf. Section~\ref{sect:synthetic data}. Then, AI-based or hybrid modeling can be used for domain specific design or implementation problems on different abstraction levels, cf. Section~\ref{sec:AI-driven} and Section~\ref{sec:HybridModeling}. In system integration or verification loops, identification, optimization and control techniques, as well as reinforcement learning methods can be applied, cf. Section~\ref{identification, optimization and control} and Section~\ref{Reinforcement Learning}.

\subsection{Supplementing experimental data by synthetic data}
\label{sect:synthetic data}
To develop robust prediction tools using machine learning (ML) methods, consistent, accurate, and rich datasets are a prerequisite. The use of ML-based methods in traditional engineering domains has been limited by the lack of datasets, since performing experiments can be time-consuming and expensive. While experimental data represents all physical effects that are present in a system, they can also suffer from noise, biases, and inconsistencies. Moreover, limited observability of all relevant system states can be an issue. Numerical models 
could overcome some limitations of experiments and provide abundant data to train ML models, potentially at the risk of introducing a domain shift. Moreover, depending on the complexity of the problem, models, e.g., FEM models, too can incur high costs due to large computational time. Generative learning-based models could overcome these limitations by training on the limited data obtained from experiments and simulations and generating realistic synthetic data. These models have gained a lot of traction in recent years due to their ability to model the actual data distribution and generate novel data points, rather than model just the decision boundary between classes. These generative models, however, have not been applied extensively in engineering domains. A few studies exist in literature that have used Generative Adversarial Networks (GANs) to supplement data in the engineering domain \cite{mctcr2023,avcrash2023,GAO2021}.

\subsection{AI-driven modeling}
\label{sec:AI-driven}
An overarching theme in supervised machine learning is the approximation of an unknown function from training data. Such a function takes an input (e.g., measurements from a production process such as temperatures or stresses) and produces an output (e.g., the statement whether the product is of high quality or not). Given data, the goal of machine learning is to train a model that is then capable of predicting the function output for previously unseen inputs. The goal of this section is to shed light on the large number of approaches and techniques that have appeared in recent years in the context of complex systems modeling.

\subsubsection{Surrogate modeling and dimensionality reduction}
\label{subsubsec:SurrogateModeling}
Similar to the above-described example, the modeling of complex systems can be cast as a function approximation problem as well. A vast number of engineering systems can be described by \emph{differential equations}, in which the derivatives of the system's state $x$ against time (\emph{ordinary differential equations (ODEs)}) or against both space and time (\emph{partial differential equations (PDEs)}) form a  functional relationship with the state itself. 
Examples for systems described by ODEs are the dynamics of multibody systems, robot movements, or molecular dynamics. PDEs, on the other hand, model distributed phenomena such as heat flow or continuum mechanics.
In ODEs and PDEs, given an initial condition $x(t_0)=x_0$, and for PDEs, appropriate boundary conditions, we would like to predict the system dynamics over some time horizon $t\in[t_0,t_e]$, while potentially considering external inputs $u(t)$.
The evolution of this state is then determined by the evolution law of a dynamical system, which can be both of continuous-time or discrete-time nature
\[\dot{x}(t)=f(x(t),u(t)) \qquad \mbox{or} \qquad x(t+\Delta t) = \Phi(x(t), u(t)), \quad x\in \mathbb{R}^{N}.\]
In the case of PDEs, where the state $x(\xi,t)$ additionally depends on a spatial coordinate $\xi$, the traditional approach is to discretize the state in space using, for instance, finite elements, to obtain a large system of coupled ODEs.
Moreover, the static case (i.e., $f(x) = 0$ or $\Phi(x) = x$, respectively) is a relevant and challenging approximation task.

There are several situations in which analytical or numerical solution techniques are infeasible: (i) the system dynamics are unknown, (ii) some parameters in an equation are unknown, or (iii) the system is simply too expensive to simulate. As a remedy, we can try to learn the system behavior from data, for which one can consider multiple approaches:
\begin{itemize}
    \item Learn the right-hand side of $f$ of the continuous-time dynamics. The resulting ML approximation can then be included in a numerical integration scheme (such as a Runge-Kutta method) to predict the system behavior.
    \item Directly approximate the discrete-time dynamics $\Phi$ (i.e., the flow map) such that the model becomes a stand-alone predictor.
    \item Noting that the solution $x(t)$ is itself a function, one can approximate $x(t)$ (or $x(\xi,t)$) directly, meaning that the time $t$ (and potentially the position $\xi$) are the inputs to the machine learning model, and $x$ is the output.
\end{itemize}
Independent of which approach is chosen, one can distinguish between \emph{intrusive} and \emph{non-intrusive} modeling, where intrusive means that the equations of motion need to be known and play a prominent role in deriving the model (for instance, via a projection step). 

\paragraph{Dimensionality reduction:} 
Prior to addressing the task of learning a predictive model, one may have to address the challenge of a highly dimensional state-space. While directly predicting the state $x$ itself is the most obvious approach in model learning, there are many situations where its dimension (i.e., the number $N$ of degrees of freedom) is prohibitively large.
Machine learning approaches such as \cite{HJE18} may be capable of overcoming this challenge by automatically identifying suitable low-dimensional structures within the high-dimensional problem. That is, we define a \emph{reduced-order state} $z$ with dimension $\ell \ll N$. Assuming that the system dynamics evolve on a lower-dimensional subspace (i.e., an \emph{attractor} or invariant manifold) -- or at least that the dominant/relevant dynamics do -- a dimensionality reduction can be learned that preserves most of the information. In this area, popular approaches are (1) the identification of linear subspaces, e.g., via the \emph{Proper Orthogonal Decomposition} \cite{Sir87} (in statistics also known as the \emph{Principal Component Analysis}), (2) the identification of nonlinear subspaces (e.g., via \emph{diffusion maps} \cite{CL06}), or (3) nonlinear coordinate transforms via \emph{autoencoders} \cite{OR19,CLKB19,PTMS22}.

\paragraph{Learning the dynamics:} 
To learn a dynamical system from data, we require time series information of the state $X = [x_0, x_2, \ldots, x_{M-1}]\in\mathbb{R}^{N\times M}$  as training data.
This data might already have undergone dimensionality reduction techniques. 
To learn discrete-time systems, we partition $X$ into two data sets $X' = [x_0, x_1, \ldots, x_{M-2}]$ and $X'' = [x_1, x_2, \ldots, x_{M-1}]$, such that $X'' = \Phi(X')$ in a column-wise fashion. Approximating $\Phi$ thus becomes a supervised learning problem.
Learning the right-hand side $f$ of a continuous-time ODE additionally requires time derivatives $\dot{X}$, which either need to be collected directly or approximated via numerical differentiation~\cite{BKB20}.
Approximating $f$ is also possible, such that $\dot{X}=f(X)$ is again a supervised learning problem. 
Denoting the approximators (parameterized by $\theta$, e.g., the trainable weights of the approximator) as $\Phi_\theta$ and $f_\theta$ , the learning problems can be formalized as
\begin{equation*}
\min_{\theta} \frac{1}{M}\sum_{m=0}^{M-1} \left\|\dot{x}_m - f_\theta(x_m)\right\|_2^2 \qquad \mbox{or} \qquad \min_{\theta} \frac{1}{M}\sum_{m=0}^{M-1} \left\|x_{m+1} - \Phi_\theta(x_m)\right\|_2^2.
\end{equation*}
Depending on the specific method, these loss functions can be augmented by additional terms for regularization, sparsification, or to take into account system knowledge such as the governing equations. For linear models, the minimization can usually be realized by linear regression (computation of the pseudo inverse), whereas nonlinear models generally require iterative procedures such as gradient descent. 

The following list presents a number of popular examples for learning dynamical system from data.
\begin{itemize}
    \item Black-box deep learning of $\Phi$ using recurrent architectures such as \emph{long-short-term memory (LSTM)} \cite{HS97,VBW+18,BPB+20}, feed-forward neural networks, or transformers \cite{GZ22}; cf.\ \cite{LZ21} for a survey.
    \item Alternatives for learning $\Phi$ using the Mori-Zwanzig formalism \cite{CHK00} or regression techniques such as ARIMA \cite{SS17}, \emph{Reservoir Computing} \cite{JH04,PHG+18}, and extreme learning machines \cite{HRV+24}.
    \item Black-box deep learning of $f$ using \emph{neural ordinary differential equations} \cite{CRBD18}.
    \item Sparse regression techniques \cite{BPK16,RBPK17}: here, the goal is to not only learn a predictor, but to identify the underlying equation that created the data. The goal is thus also the identification of potentially unknown systems or their parameters.
    \item Identification of linear systems / dynamic mode decomposition (DMD) / Koopman operator: 
    For control and trajectory optimization purposes, a particularly useful class of surrogate models can be created by lifting a nonlinear dynamical system into a higher dimensional state space.
    In this lifted space, the dynamics can be represented as linear, for example, using the Koopman operator \cite{Sch10,RMB+09,KNP+20,Mauroy2020,BBKK22}. 
    This (partially) resolves inherent non-convexity of the otherwise nonlinear optimization problems and can thus aid with the design optimization of dynamical systems. 
    In particular, gradient information becomes easily accessible, which is crucial in optimization problems with a high-dimensional search space.
    The lifted dynamics can be identified from data through an \emph{extended} dynamic mode decomposition (EDMD), which extends DMD by including a large library of observables; that is, of lifted states \cite{Williams2015}.
    This concept can be extended to representations that facilitate control, either in the form of a linear time invariant system \cite{PBK16,KM18}, as a switched system \cite{PK19}, or as a bilinear representation with linear autonomous dynamics and a bilinear control input \cite{POR20,Bruder2021}. 
    \item Intrusive modeling of PDEs via Galerkin projection \cite{Sir87,KV02,BGW15,POD19}; this type of surrogate modeling has been extremely popular already before the machine learning revolution. The underlying approach is, to learn an $\ell$-dimensional linear subspace $\Psi=\{\psi_1,\ldots,\psi_\ell\}$ of $X$ with the smallest possible projection error, which we obtain by performing a singular value decomposition (or principle component analysis, which is the same). Inserting a Galerkin ansatz $x(t,\xi) \approx \sum_{j=1}^\ell a_{j}(t)\psi_j(\xi)$ into the weak form of the PDE then yields an $\ell$-dimensional ODE for the dynamics of the coefficients $a$.
    \item Physics-informed machine learning \cite{RPK19,KKLP+21,LKA+21}: The task of physics-informed neural networks (PINNs) is the direct approximation of the mapping from a given input time $t$ (and potentially position $\xi$) to the corresponding system state $x(t)$ (or $x(t,\xi)$). When approximating this mapping by a neural network, one can easily calculate derivatives of the output $x$ with respect to the inputs $t$ (and $\xi$) -- simply using numerical differentiation/backpropagation. If we know the differential equation of the system we would like to predict, then we can simply construct a loss function that encodes the physics.\\
    Consider the heat equation $x_t - \lambda x_{\xi \xi}=0$, for example. In this case, one could train the neural network weights $\theta$ in such a way that the model output $x^\theta$, differentiated with respect to the input variable $t$, minus $\lambda$ times $x^\theta$ differentiated twice with respect to the input variable $\xi$, is minimized: $\min_\theta \sum_{m=0}^{M-1} \| x^\theta_t (t_m, \xi_m) - \lambda x^\theta_{\xi \xi}(t_m, \xi_m)\|_2^2$, where the loss is evaluated on $M$ collocation points $(t_m, \xi_m)$ distributed in space and time.
\end{itemize}

\paragraph{Generative modeling:} Generative learning-based models, since the advent of Generative Adversarial Networks (GANs) \cite{goodfellow14}, have gained a lot of traction in the field of AI. GANs \cite{stylegan,3dgan} and Diffusion models \cite{diffusion,diffusion2} have demonstrated remarkable progress in high-resolution image and video generation and large language models (LLMs) have become household names due to the success of AI-based text-generators such as ChatGPT \cite{chatgpt}, Gemini \cite{gemini}, and Llama \cite{llama} for their ability to generate near humanlike responses. Generative models such as GANs have tremendous potential for surrogate modeling, especially modeling FE simulations. Since GANs do not map inputs to outputs and rather learn from the underlying patterns in data, a well-trained GAN can overcome certain problems pertaining to supervised learning, such as overfitting, accurate interpolation inside the data range, and extrapolation of results \cite{rpamm2023}. The very first publications in this area \cite{siddani21,li23,cheng20,rgan2024} explore a data-driven approach in predicting the field variables in a finite element simulation.

\subsubsection{Application domains} \label{sec:application_domains}
To illustrate the application of these techniques, the following section provides a number of examples, motivated by the research performed in the priority program SPP 2353.
We not only highlight the potential for the use of AI-based modeling and optimization techniques, but also elaborate on application-specific challenges.

\paragraph{Mechanical engineering:} 
Designing mechanical mechanisms is a challenging task, since linkages and other moving structures are subject to complex physical constraints and often exhibit discrete design choices. 
Additionally, the relationship between the motion and the mechanism design can be highly non-linear especially at the boundary between different mechanism classes.
As a consequence, multiple --potentially topologically disconnected-- solutions may exist for a given task.
Nevertheless, carefully designed special-purpose machines can potentially be lighter, more energy-efficient, easier to control, and more precise compared to general-purpose systems.
Hence, extensive research has been carried out using analytical formulations or optimization-based approaches~\cite{ullah1997,cabrera2002,liu2004,smaili2005,sedlaczek2009,buskiewicz2010,kang2016,huang2021,kang2022,nguyen2022,baskar2023}.
Analytic approaches require profound mechanism knowledge, while optimization relies on sequential simulations, which have to be performed repeatedly for each new task.
In contrast, data-driven methods require an extensive training stage but afterwards can almost instantaneously and with little calculation effort propose new designs.
A review of different deep learning approaches for a broad range of engineering tasks, including a section on kinematic synthesis, is provided in~\cite{regenwetter_deep_2022}.
In \cite{nobari_links_2022}, a data set of 100 million different linkages is presented to foster the development of data-driven methods and to allow for benchmarking.
In~\cite{yim_big_2021}, a neural network is used to predict the discrete coefficients of a spring-block model for the design synthesis of planar mechanisms, followed by a gradient-based local optimization for the dimensional synthesis. The method has been extended to also allow for the synthesis of spatial linkage mechanisms~\cite{yim_big_2023}.
In one author's group at the ITM investigated different feature extraction methods and normalization strategies for the representation of paths used to predict four-bar linkage parameters~\cite{roder_towards_2023} .
To cope with the multi-modality of possible solutions, individual deep neural networks can be trained on subspace regions of the design space~\cite{chen_application_2023}.
An alternative approach is presented in \cite{lee_deep_2024}, which proposes the usage of conditional generative adversarial networks for the synthesis of crank-rocker mechanisms, including kinematic and quasi-static conditions. 
The conditioning of the generator can then be used to propose different mechanism designs even, for the same target conditions.
Apart from the design synthesis of mechanisms, AI methods can also be utilized for tasks that have previously been impossible.
Natural-language models can be used to create multibody simulation code~\cite{gerstmayr_multibody_2024} or to constitute multibody models from hand-drawings of mechanisms for rapid prototyping~\cite{nurizada_transforming_2023}.
Finally, \cite{deshpande_machine_2019, deshpande_image-based_2020, nurizada_invariant_2023} use autoencoder networks to represent target paths in a compressed lower-dimensional feature space.

\paragraph{Robotics and Mechatronics:} 
In the design of robotic systems, desirable performance is often characterized by speed and energetic efficiency. 
With these goals in mind, the design of robotic systems can be challenging, as the systems' hardware (i.e., the morphology) and the associated control (i.e., the motion) are inherently coupled via the mechanical dynamics.
Consequently, a co-design approach must be applied to simultaneously optimize both motion and morphology~\cite{Spielberg2017,Ha2018,Yesilevskiy2018,Fadini2021,BravoPalacios2022}. 
If a diverse set of tasks is desired, this approach must trade-off a single mechanical design to optimally support different motions that the robot needs to perform efficiently.
Such multi-task optimizations can lead to high-dimensional, coupled, and complex optimization problems.
Solving these problems has traditionally required significant human intervention and expert knowledge to identify suitable robotic designs. 
More recently, AI has become an important element in the design processes for robotic systems. The applications of machine learning techniques span from genetic algorithms~\cite{Mombaur2005,Geijtenbeek2013,Coros2011} to more recent advancements in deep reinforcement learning~\cite{Schaff2019,Gupta2021,BelmonteBaeza2022}.
To aid with traditional, optimization-based approaches to co-design, AI-based modeling techniques, as outlined above, could help alleviate the computational burden. 
Here, models that facilitate the evaluation of sensitivity information as they are needed in optimization and control are favored.

\paragraph{Acoustics:} 
Machine learning techniques are gaining momentum in different areas of acoustics. Applications range from ocean acoustics \cite{choi2019acoustic}, speech processing \cite{xu2014regression}, environmental acoustic monitoring \cite{gontier2021polyphonic} to musical acoustics \cite{jourjine2000blind}. A general overview for machine learning in acoustics was assembled in 2019 \cite{bianco2019machine}. A review specifically dedicated to machine learning for vibroacoustics is provided in \cite{cunha2023review}. In the following, we focus on ML applications related to product design, in particular, we discuss achievements in the field of room acoustics, metamaterial design and fault detection. \\
Room acoustics aim to analyze acoustic properties, such as the mean absorption coefficient, reverberation time or speech quality of a room. In \cite{de2021unsupervised}, ML techniques are investigated to identify human and mechanical noise sources in noisy office environments. To this end, Gaussian mixture models and K-means clustering techniques are used to identify the respective sound sources. To analyze the acoustic characteristics of an existing room, \cite{foy2021mean} proposed supervised ML to predict the mean absorption coefficients based on the room impulse response. The work \cite{xiao2015learning} targets direction and arrival estimation in noisy and reverberant environments. A supervised learning approach is presented for robust direction of arrival estimation based on simulated microphone array inputs.  

Acoustic metamaterials are periodically arranged structures, which promise advantageous acoustic properties, e.g., high sound insulation effects. To predict the sound absorption coefficient, \cite{ciaburro2021modeling} trained a supervised learning model on measurement data and demonstrated good generalization properties. To tackle the challenge of metamaterial design, \cite{gurbuz2021generative} proposed conditional generative adversarial networks to create new metamaterial design proposals. New metamaterials are designed, which follow a desired transmission behavior in the frequency domain. In \cite{bacigalupo2020machine}, Radial Basis Function networks and Quasi-Monte Carlo methods are used for surrogate based optimization of transmission and dispersion properties of acoustic meta materials.

Non-destructive testing is an important technique for maintaining products during lifetime. The work in~\cite{shafiei2021using} investigates ultrasonic inspection of polyethylene pipes, which are used for gas distribution. ML techniques are used to detect failures, in particular convolutional neural networks (CNNs) are successfully applied in a four-class classification problem. In the field of robot-assisted ultrasonic testing, \cite{mei2021visual} applied ML techniques to improve the image resolution. To this end, a visual geometry group U-Net processes the ultrasonic images.

\paragraph{Fluid mechanics:} 
System modeling and design in fluid dynamics is a highly challenging task, one of the key challenges being the complex flow physics governed by the Navier--Stokes equations in various forms (incompressible/compressible, reacting/non-reacting etc., depending on the specific application). In the regimes that are relevant for applications, the flows are usually turbulent. This means that we have to deal with complex multi-scale physics in both space and time, and computations quickly exceed modern computing capacities even on high-performance computers.

Due to this, AI has become an important factor in design processes for systems governed by fluid mechanics. The usage of machine learning in fluid mechanics \cite{BNK20} can be divided into multiple categories. The equations of motion are well known, but usually too expensive to solve numerically, which is why one of the key goals is to accelerate simulations. This can be achieved using surrogate modeling in various forms laid out in Section~\ref{subsubsec:SurrogateModeling}, cf., e.g., \cite{Sir87,KV02,BWG08,PS20,PTMS22,ETSV22}.
In a similar fashion, turbulence modeling \cite{ZD15,BK22} can be used to allow for coarser mesh resolutions in space, which also facilitates faster simulations without losing accuracy.
The task of super resolution is used as a post-processing step, where convolutional neural networks are used to increase the level of detail of coarsely resolved flow field data \cite{FFT19,GSD22}.

Beside these acceleration tasks, AI and machine learning can be of great help for system understanding (e.g., in the form of modal analysis \cite{TBB+17,Sch10,RMB+09}), as well as in the design process itself \cite{BKM+21}. Two particularly popular examples are the choice of suitable sensor placements \cite{Wil06,MBKB18} and the data-driven design of controllers \cite{GAD+15,POD19,BPB+20,GRS+23,PSC+24}.

\subsection{Hybrid modeling} \label{sec:HybridModeling}
While AI-driven modeling generates a model based solely on measurements and without system knowledge, hybrid modeling incorporates existing knowledge. A hybrid model generally involves the combination of multiple submodels from various modeling methodologies, such as physical, statistical, and machine learning models, allowing the limitations of individual models to be compensated \cite{Halfmann2003, Stosch2014, Thompson1994}. For instance, the physical model of a system can serve as the foundational framework of the modeling process, which is then augmented with suitable data-driven components \cite{Psichogios1992}.
\paragraph{Phenomenological effects:}
The precise representation of the dynamic behavior of technical systems using physical modeling approaches is often challenging, especially when the involved effects can only be captured with considerable effort \cite{Stosch2014}. This includes, for instance, tribological effects, which represent the dissipative aspects of dynamic system behavior \cite{Popov}, or the complex aerodynamic forces acting on a moving vehicle \cite{Popp}. Detailed modeling approaches typically exhibit high complexity and application specificity. In situations where physical models are unavailable, impractical to implement, or complex to parameterize, data-driven models based on measurements can provide a viable alternative. Combining both approaches into hybrid modeling of dynamic systems contributes to increasing the efficiency of model development. One approach is to compensate for discrepancies between the model and measurements resulting from incomplete physical modeling by employing a data-driven model \cite{Kaheman2019, Wohlleben2022, Chen}. The differential equation to be solved can be expressed as
\begin{equation*}
    \dot{x}(t) = f\left(x(t), u(t)\right) + h\left(x(t), u(t) \right) ,
\end{equation*}
where $f$ represents the physical model and $h$ represents the data-driven model \cite{Wohlleben2024, Ebers, Farchi}. The data-driven component allows for the integration of phenomena that the physical model alone might miss.

\paragraph{Model reduction:}
Large systems consisting of multiple components and submodules pose a challenge. They often have a large number of degrees of freedom, making them computationally intensive \cite{Halfmann2003}. By replacing entire components with data-driven models, which often have easily computable structures, the complexity of the system model can be reduced \cite{Reichstein}. When employing data-driven models to replace complex components, it is essential to ensure that the internal dependencies and suitable interfaces between different parts of the system are maintained. This involves careful integration of data-driven models to ensure they interact correctly with the remaining physical models. Model reduction not only enhances computational efficiency but also aids in the design and optimization process. Simplified models allow for quicker iterations and more extensive exploration of the design space, which is particularly beneficial in multi-objective optimization scenarios. However, it is crucial to validate the reduced models against high-fidelity simulations or experimental data to ensure their reliability and accuracy. Furthermore, the modular nature of hybrid modeling supports scalability and flexibility. As new data becomes available or as system requirements change, individual data-driven components can be updated or replaced without necessitating a complete overhaul of the entire model. This adaptability is vital for maintaining the model's relevance and accuracy over time.

\paragraph{Error reduction:}
Hybrid modeling enhances the accuracy of models by leveraging the strengths of both physical and data-driven approaches \cite{Halfmann2003, Stosch2014, Psichogios1992}. The physical model provides a grounded understanding of the system's fundamental behaviors, while the data-driven component can correct for inaccuracies and fill gaps left by the physical model or by error-prone measurements \cite{Ebers, Farchi}. This synergy allows for ongoing error reduction as more data becomes available, continuously refining the model's predictive capabilities. However, the generalization capabilities of the data-driven model determine the validity regimes of the hybrid model overall.

\paragraph{Transparency:}
Hybrid modeling promotes transparency by maintaining a clear distinction between the physically understood parts of the model  and the data-driven components \cite{Lindskog, Miller}. This separation allows for better interpretability of the model's behavior, making it easier to diagnose issues, understand the contributions of different parts of the model, and communicate the model's workings to stakeholders \cite{Wohlleben2023}. Moreover, this transparency is crucial for validating and verifying the model. By clearly distinguishing between the physical and data-driven parts, it becomes easier to trace errors and understand the sources of inaccuracies. This clarity also facilitates the debugging process, as engineers can pinpoint whether issues arise from the physical assumptions or the data-driven adjustments. Transparency also enhances the model's flexibility. When the contributions of physical and data-driven components are clear, it becomes easier to update or replace parts of the model in response to new data or changing requirements. This modular approach allows for more straightforward adjustments and scalability, ensuring the model remains relevant and accurate over time.

\subsection{AI-assisted identification, optimization and control}
\label{identification, optimization and control}
In an industrial context, design is typically based on expertise and driven by a trial \& error procedure. Small changes of one or few design parameters are analyzed with often time-consuming commercial off-the-shelf (CoTS) software or multi-disciplinary analysis processes in order to gain some knowledge about improving and worsening design changes. In rare cases, this is supported by gradient information obtained from numerical differentiation or adjoint methods. In high-dimensional design spaces, however, such a procedure is rather time-consuming and will rarely lead to a truly optimal design.
Substitution by numerical multi-criterion optimization is, therefore, the better solution. Especially population-based algorithms, such as multi-criterion genetic algorithms or particle swarm optimization are able to obtain optimal tradeoffs even in case of multi-modal objective functions and heterogeneous design spaces, see, e.g., \cite{RTS01,HSG+05}. 

Since in industrial practice, design intention is mostly intuitive and only rarely formalized, one of the first steps is to find mathematical functions for the objectives and constraints. Secondly, design parametrization is mostly analysis-oriented and thereby often less qualified for optimization. In contrast to human designers, numerical optimization can only account for design restrictions formally included in the optimization problem. In shape optimization, e.g., a re-parametrization as splines (piecewise polynomial curves) may help to account for smoothness of curves and surfaces with few parameters only. In order to resolve the conflict between the requirement of tens of thousands of design evaluations for genetic algorithms and the high computational effort for analysis, surrogates may substitute direct design evaluation during optimization \cite{PD18,CJM+18,DRH20}. Machine learning models such as multi-layer perceptrons, regression trees or radial basis function networks can often be adaptively improved using a comparatively small number of expensive simulations of the original system \cite{BPB+20,BP21}. After multi-criterion optimization has found an Edgeworth-Pareto-set of multiple optimal tradeoffs, the designer may assess these designs with respect to physics and apply priority measures to select a specific one for realization.

Optimization is also the usual strategy for parameter identification. Machine learning, however, can be an alternative since it can find correlations between data in any direction. Thus, a data set of function values or time-series may be generated by classical analysis of a random set of parameters to be identified. By reversing the input-output order, machine learning may then learn a multi-layer perceptron relating parameter values to any provided function value or trajectory. In the latter case, convolutional networks are recommended to account for the high correlation between neighboring time points. Such a procedure may also be applied to control problems requiring inverse kinematics and dynamics models. In tracking problems, the obtained surrogates may then be used as feedforward controllers and accomplished by rather simple traditional feedback control.

\subsection{Reinforcement learning as an optimization method}
\label{Reinforcement Learning}
Due to the advancements made to reinforcement learning (RL), it is gaining popularity in various domains, such as engineering \cite{matallah_2022_rl_powertrain, pandit_2024_DRL_Silica, Qin_2021_cascade_blade}, robotics \cite{Tao_2023_Jumping_bipedal, sehgal_2019_DRL_genetic}, economics \cite{Zheng_2020_AI_economist}, data management and planning-control \cite{klar_2022_scalability} along with computer science problems \cite{mnih_2015_human, Gu_2016_continuiusRL}. In RL, agents operate in an environment (or a working domain) and take specific actions to change the state of the environment. The agents receive feedback in terms of a reward signal from the environment and then learn through sampling to take better actions to maximize their cumulative reward \cite{sutton_2018_rl}. The RL agents can learn continuously and improve as they interact with the environment. Therefore, RL can be used in parameter exploration, optimization, function approximation, automation, and control problems. It has been shown that RL agents can learn to perform as well as or better than humans in various gaming environments with the introduction of the base deep Q-network (DQN) agent \cite{mnih_2015_human}. Furthermore, multiple user-friendly libraries exist with these RL agents, which can be directly implemented to understand and solve complex problems like multiobjective inverse optimization. 

In inverse optimization problems, parameters are optimized to fulfill precise objectives while considering various physical and non-physical constraints, traversing through parameter space, and exploring the feasible parameters. Evolutionary algorithms (EA) are conventionally used to solve the optimization problem. Still, they have a fundamental drawback in terms of the stochastic character of action and its effect, which can cause delay. In some large-scale problems, EA cannot be applied without a large population size or a number of iterations \cite{schmidhuber_2000_EA_DRL}. Therefore, deep reinforcement learning (DRL) methods are explored \cite{schmidhuber_2000_EA_DRL}. Complex optimization problems can be modeled as a model-free environment, and the optimizing policy can be evaluated and adjusted. The agent can sample from the custom optimization environment and explore the parameter space. The reward can be used to guide the agent in fulfilling the objectives while exploring. Therefore, these DRL methods are increasingly used with EA methods in various optimization problems \cite{Bai_2023_EARL, Chen_2020_GARL, DRUGAN_2019_RL_Survey}.

\section{Examples of the integration of AI into the design process}
\label{sec:example}

In this section, we present illustrative examples from the SPP~2353\footnoteref{note1} initiative that demonstrate the innovative use of AI in the design development of new systems. These examples show how AI technologies are integrated into the design process to improve efficiency, creativity, and the overall design outcome.\\
Each example begins with a brief problem statement to provide the context. We then address four key questions to illustrate the role and impact of AI in the design process:
\begin{enumerate}
\item What \textbf{prior knowledge or experience} can be drawn upon to inform the design process?
\item What \textbf{models} exist and are used to improve the design outcome?
\item What \textbf{limits} exist in the design process to date and what \textbf{potential} does the use of AI methods harbor?
\item What \textbf{data} has been collected and processed for the design process?
\end{enumerate}
Finally, we give an outlook on the expected results and offer insights into the expected progress and benefits of integrating AI into the design process. With these examples, we aim to illustrate the transformative potential of AI in developing innovative design solutions and overcoming traditional challenges.

\subsection{AI based design assistance system for soft robotics – optimizing complex systems based on the smallest design entity (Project of K.M. de Payrebrune)} \label{sec:SoftRobotics}
\begin{figure}[hbt]
\centering
\includegraphics[width = .9 \textwidth]{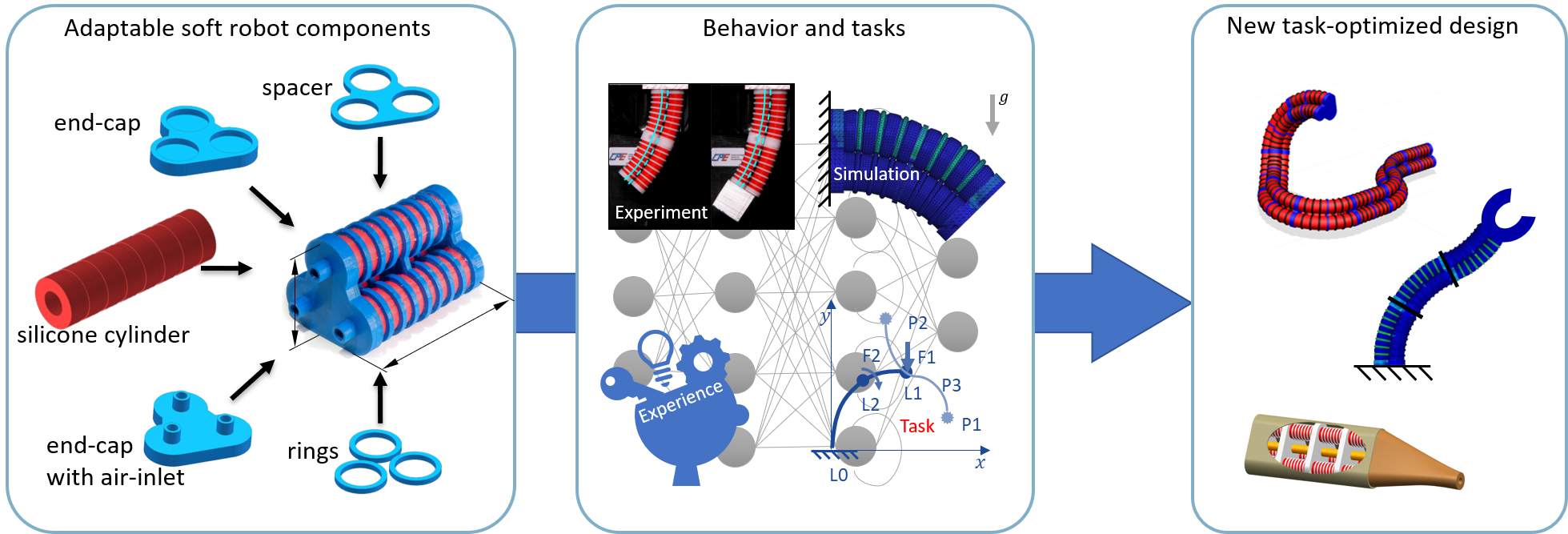}
\caption{AI-design assistance for soft robots with task-optimized components.}
\label{fig:softrobot}
\end{figure}

The main goal of this project is to develop customized soft robots that are designed to perform specific tasks. The robots have a modular design, whereby the dimensions of each module can be individually adapted. 
To achieve an optimal design of a soft robot consisting of several modules, we use machine learning algorithms to improve model extensions, to increase the prediction accuracy of robot performance, and to determine the optimal design based on the knowledge of one module, as outlined in Fig.~\ref{fig:softrobot}. 

\begin{enumerate}
\item \textbf{Prior knowledge/experiences:} The development and design of soft robots is usually based on the creativity of the researcher and the research questions that are to be investigated. The improvement in performance is usually achieved through trial \& error. In particular, the complex interaction between the chosen structure and applied loads makes it difficult to predict the continuous deformation of the robot.  In addition, the design of the robot must be adapted to the manufacturing possibilities. As the robots are usually made of silicone and controlled with compressed air, additional restrictions exist here. 
\item \textbf{Model:} Various model approaches exist for the calculation of soft, slender structures with continuous deformation. Among the beam models that reduce the three-dimensional structure of the robot to a material line in space, there are continuum mechanics-based approaches with constant curvature per robot section \cite{Webster2010}, Cosserat beam approaches with variable curvature \cite{Eugster2022} and tendon driven kinematics \cite{Till2019}. In parallel, finite element formulations of varying complexity (shells \cite{renda2018unified}, isogeometric shells \cite{liu2020isogeometric}, three-dimensional description \cite{Runge2017b}) are used. First model-free approaches use data-driven machine learning and deep learning algorithms to find a relation between input and output \cite{Schindler2024a}. Particular challenges lie in the dynamic description, as well as in their control, as the robots react sensitively to external disturbances and loads.
\item \textbf{Limits and boundaries:} In our project, we aim to enhance the prediction and design of complex soft robot structures by leveraging machine learning algorithms, especially by hybrid models incorporating data-based sub-models. Firstly, we seek to expand the beam model by incorporating influences such as nonlinearities in material behavior and friction between the silicone body and other components of the soft robot module. These factors are challenging to integrate into the existing physical model or have not been addressed adequately at all. To achieve the model adaptation, simulation results will be compared with measurements of soft robot deformations, and model parameters will be iteratively adapted.
Secondly, based on 3D finite element simulations and solutions derived from beam formulations, we aim to establish a relationship between design variables and deformations for various pressure actuations and loads. This will enable predictions of performance for further design modifications, facilitating a more precise adaptation of the soft robot design to specific tasks.
\item \textbf{Data:} In order to read out the deformation of real soft robot modules for variable pressurization, photos are taken with a stereo camera system and evaluated with a modified iterative close-point algorithm for marker-free 3D shape reconstruction \cite{Schindler2024a, Schindler2024b} from the group of Sattel and Flaßkamp \cite{Hoffmann2024}. 
The backbone curves obtained can then be used with simulation results to adjust the model parameters. 
\end{enumerate}
We expect to develop a design assistant tool that helps to find an optimal design solution of a soft robot for a specific task. The design assistant tool will combine results from machine learning approaches and experiences of the user such that additional aspects, as the ability to manufacture the robot, can be included in the design process. Based on the numerical analyses, a faster and more reliable design of a new soft robot system will be achieved. 

\subsection{AI-design assistant for stereotactic neurosurgery with concentric-tube continuum robots (Project of T.~Sattel and K.~Flaßkamp)}\label{sec:Neurosurgery}

\begin{figure}[h]
\centering
\includegraphics[width = .8 \textwidth]{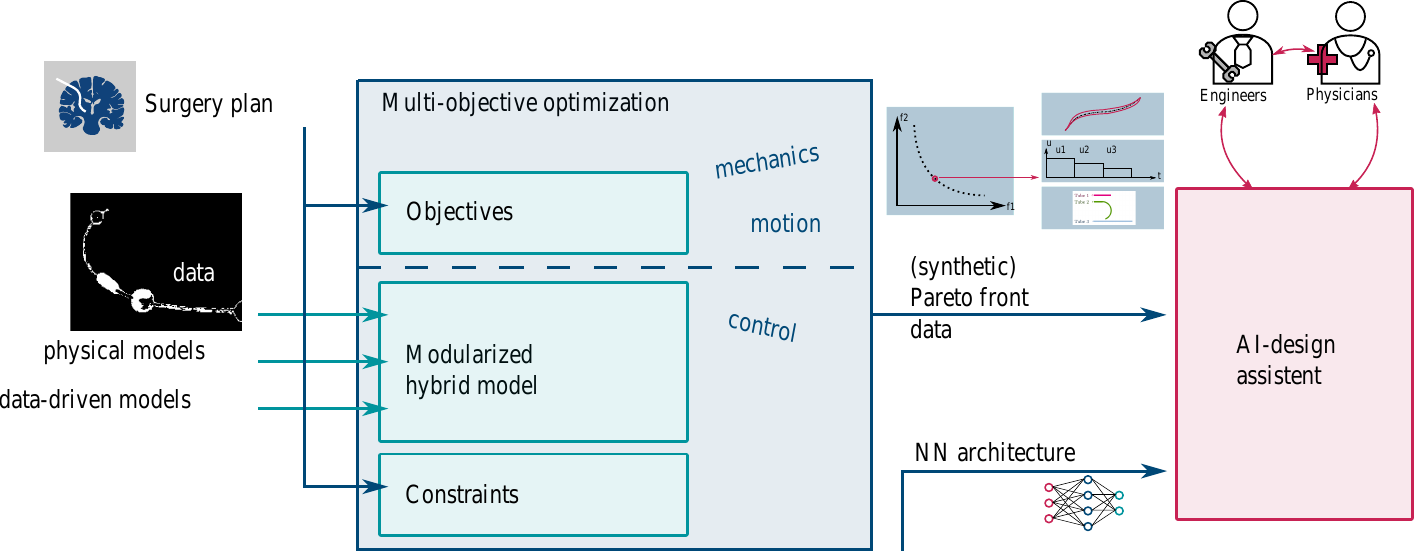}
\caption{AI-design assistance for stereotactic neurosurgery with curved cannulae.}
\label{fig:FlasskampSattel}
\end{figure}

The aim of this project is to develop patient-specific tools for stereotactic neurosurgery. The tools are on the one hand hardware tools, i.e.,~surgical tools here named as curved cannulae, configured as nested NiTi-tubes, and second software tools, i.e.,~methods and algorithms to plan the optimal surgery, which keeps brain damage to a minimum, see Fig.~\ref{fig:FlasskampSattel}. We combine first-principle models with data-based modeling approaches and solve the coupled design-planning problem by combinations of gradient-based optimization, evolutionary algorithms, and learning methods. Design parameters for the cannulae are the geometrical data of the tubes and their pre-curvature. Design objectives are minimal path length, target point accuracy, distance to critical regions, etc. Restrictions are e.g.~maximal curvature or geometric limitations. The design output are a bunch of cannulae configurations and assigned control parameters for steering them via translation and rotation of their proximal ends safely to the target point in the brain.

\begin{enumerate}
\item \textbf{Prior knowledge/experiences:} It is state-of-the-art to perform minimally invasive deep-brain stimulation or biopsies by stereotactic neurosurgery with a single straight cannulae, \cite{Muehlenhoff2022}. Curved cannulas offer more flexibility in reaching the target with potentially less risk. Concentric-Tube Continuum Robots (CTCR) are the underlying technological base for the application of curved cannulae in minimally invasive surgical procedures, \cite{Burgner-Kahrs2015}. But curved cannulae need sophisticated planning and assistance tools, not available yet. Physicians' experience is needed for annotation of MRT data to define the constraints (i.e.\ vessels and sensitive brain areas) for the surgery path to be planned within the brain, to identify possible entry points on the skull and to mark the target points.
\item \textbf{Model:} Detailed mechanical models for nested cannulae configurations have been developed in previous works, e.g.~\cite{Rucker2011,Greiner-Petter2017}.
Parameter identification is needed based on experimental data.
Some material and geometrical effects (non-concentricity of tubes; nonlinearities, in general; elasticities, hysteresis) are not included or only partially included, thus the model accuracy is not fully satisfactory.\\
Defining typical cost functions such as "shortest path to target" and modeling constraints given by the brain topology in a numerically processible way, e.g., by ellipsoids, we can solve the planning problem by optimal control methods, e.g.~\cite{HoffmannEsterhuizen2023,SAUERTEIG2022600}.
Parametrized cannuale models add a few more parameters to the optimization problem for finding simultaneously an optimal path with an optimal cannula configuration.
\item \textbf{Limits and boundaries:}
The model accuracy shall be improved by \emph{hybrid models} including data-based submodels.
We have used \emph{iterative learning control} to improve tool-tip accuracy between photogrammetric data and model-based simulations.
Since production of tubes is costly and time-consuming, real data of different cannulae configurations is rare. This raises a need for artificial data of possible cannulae design which can be evaluated regarding their performance in surgery planning.
Optimal path planning with classical gradient-based methods is computationally too demanding, in particular in multi-objective problems. In fact, the aim is to learn optimal solutions of parametrized problems.
\item \textbf{Data:}
Nested-tube cannulae configurations have been tested in a concentric tube continuum robot in a \emph{photogrammetry system}, \cite{Muehlenhoff2022}.
\emph{Simulation data} can be generated by Cosserat beam models.
\emph{Artificial data} has been generated by GANs in collaboration with M.~Stoffel et al., \cite{mctcr2023}.
\end{enumerate}
We expect to develop AI-assisted planning tools for patient-individual stereotactic neurosurgery, including design optimization and path planning for nested NiTi-cannula configurations.

\subsection{AI-design assistant for simplifying optimization of complex technical systems (Project of P. Eberhard and D.~Bestle)}\label{sec:BestleEberhard}
\label{sec:Optimizationtechnicalsystems}

The project aims at reducing the required expert knowledge about an underlying system to be designed by using numerical optimization strategies instead of performing expert-based design improvements.

\begin{enumerate}
\item \textbf{Prior knowledge/experiences:} In industrial applications, usually technical systems are still designed by trial \& error. Typical arguments against using numerical optimization strategies are high computational analysis times, that classical optimization algorithms propose rather impractical designs, or missing gradients required by efficient optimization algorithms when integrating CoTS software for analysis.

\item \textbf{Model:} Design assistants integrating modern optimization concepts for multicriterion optimization may offer an Edgeworth-Pareto-set of several optimal tradeoffs where the design engineer may choose from while releasing him from tedious search. Genetic algorithms may find these tradeoffs even in case of multimodal objective functions where the requirement of tens of thousands of design evaluations is avoided by using surrogates derived from few original design analyses by machine learning. Hidden constraints resulting in failed analyses may be represented by classification trees or neural classification networks to predict infeasible designs and thereby avoiding unnecessary analyses.

\item \textbf{Limits and boundaries:} Both surrogate training and optimization with evolutionary algorithms suffer from the curse of dimensionality. Statistic sensitivity analysis like e.g., computation of Sobol indices may provide information, which allows reducing the design space dimension to the most influential parameters. Also here surrogates may provide an artificial big data set required for computing these indices.

\item \textbf{Data:} The data as basis for surrogate training may be obtained from simulation or analysis of models derived from first principles of physics or mechanics like multibody system formalisms. By using an adaptive concept, where approximate surrogates are first trained with small data sets obtained from design of experiment strategies like Latin hypercube sampling and then refined only in promising design regions proposed by the optimization algorithm, may cut down the number of original design evaluations dramatically and thus makes automated design feasible even for complex systems.
\end{enumerate}

\begin{figure}[h]
    \centering
    \includegraphics[scale=1.1]{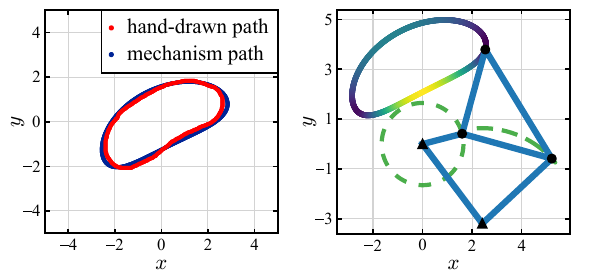}
    \caption{The automatic design of a multibody system for a user-defined path.}
    \label{fig:multibodydesign}
\end{figure}

The project will develop above concepts along the final goal to design controlled flexible multibody systems in order to show their applicability. An example is shown in Fig.~\ref{fig:multibodydesign}.

\subsection{Data-driven machine learning enhanced optimization of vehicle crashworthiness design (Project of M. Stoffel)} \label{sec:vehiclecrashopt}

\begin{figure}[h]
    \centering
    \includegraphics[width=0.9\textwidth]{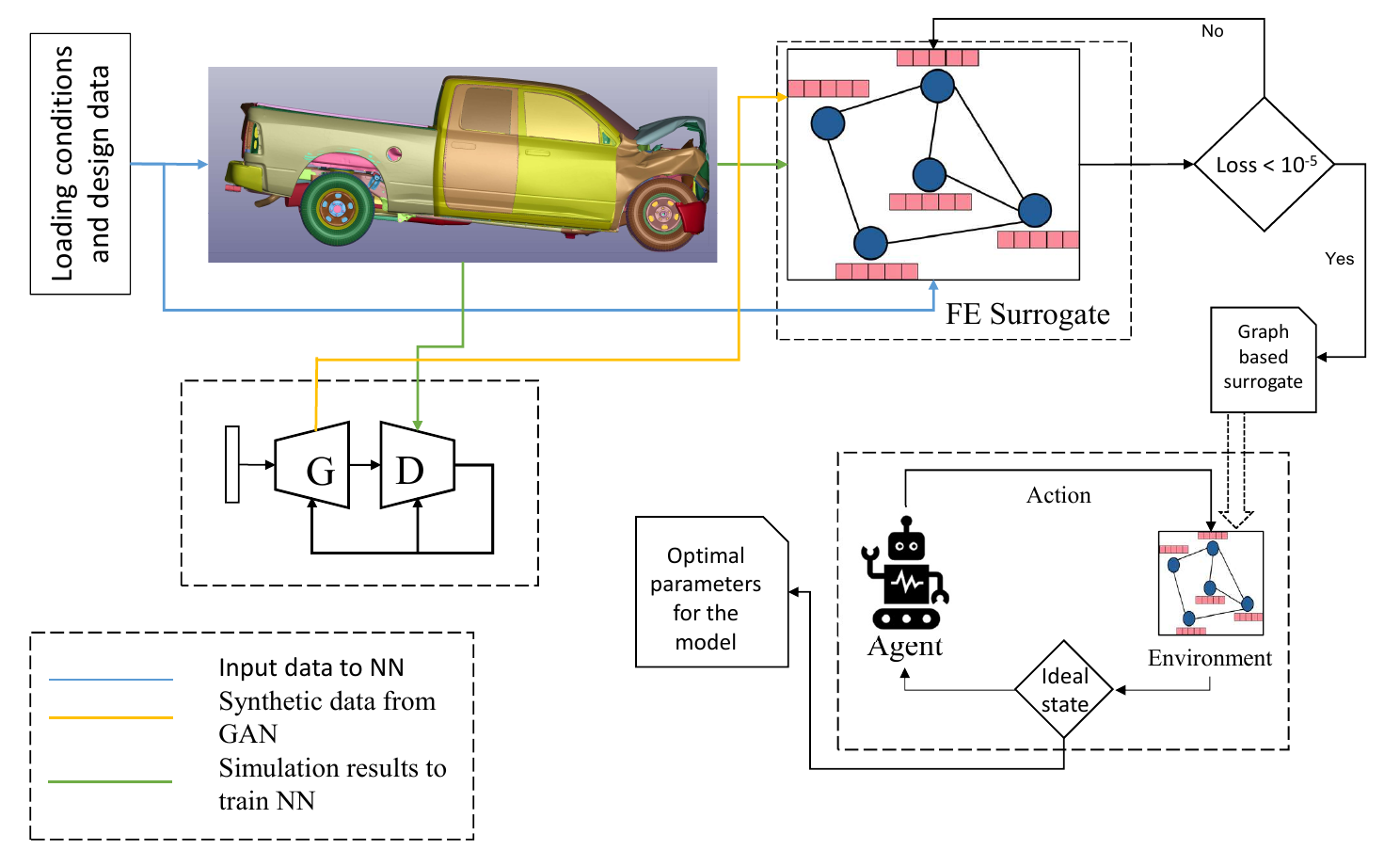}
    \caption{Design assistant framework for optimization of crashworthiness.}
    \label{fig:crashFrameWork}
\end{figure}

The main goal of the project is to develop a design optimization framework based on Reinforcement Learning (RL) with the ability to supplement with synthetic data from generative learning models. The complete framework, as shown in Fig. \ref{fig:crashFrameWork}, has been applied to vehicle crashworthiness design here but can be extended to any design optimization task. Generative modeling is utilized to generate synthetic data to supplement the numerical data to make the machine-learning models more robust. 

\begin{enumerate}
    \item \textbf{Prior knowledge/experiences:} Vehicle design requires complex systems engineering, where a large number of attributes must be systematically considered. Most of the current commercial optimization tools offer various methods for mathematical surrogates. These are also known as meta-models and regression models and are used in response surface methods to reduce the computational effort required for the design optimization process. The second widely used approach to saving computational time is employing physical surrogates, simplified or reduced computational models realized from sub-structure modeling, reducing the model to rigid bodies, rough-fine meshing, or making equivalent multi-body models. 
    \item \textbf{Model:} Several classical approaches based on evolutionary algorithms and heuristic methods exist for the optimization of vehicle crashworthiness design \cite{fang2017design}. These models, however, are time-consuming, adequately accurate, and do not work for multiparameter and multiobjective optimization tasks. Expert input is paramount for using these methods to set up the design of experiments (DOE) and choose the best possible result. 
    A RL-based framework is proposed, that can handle multiobjective and multiparameter optimization tasks with relative ease and eliminate the requirement of an expert input in the optimization process.

    \item \textbf{Limits and boundaries:} Training generative learning-based models is challenging due to the indirect nature of their training, and they are susceptible to hyperparameters. Since non-linear regression with generative models is a novel approach, customized loss functions must be used in addition to the likelihood principle. These functions cannot be generalized. 
    \item \textbf{Data:} Impact simulations for passive safety components of vehicles have been performed using Abaqus and LS-Dyna. Since these simulations consume a lot of computational time and effort, a crashbox is used as an example model to demonstrate the optimization method. These datasets have been further enriched with synthetic data from generative models and used to train the RL model.
\end{enumerate}
The project develops an overall framework for design optimization and presents multiple design assistants, such as the generative model, FE surrogate, and RL model, which can be independently utilized.

\subsection{Hybrid modeling for data-enhanced multiobjective optimization of multibody systems (Project of W. Sextro and S. Peitz)} \label{sec:HyM3}
\begin{figure}[h!]
    \centering
    \includegraphics{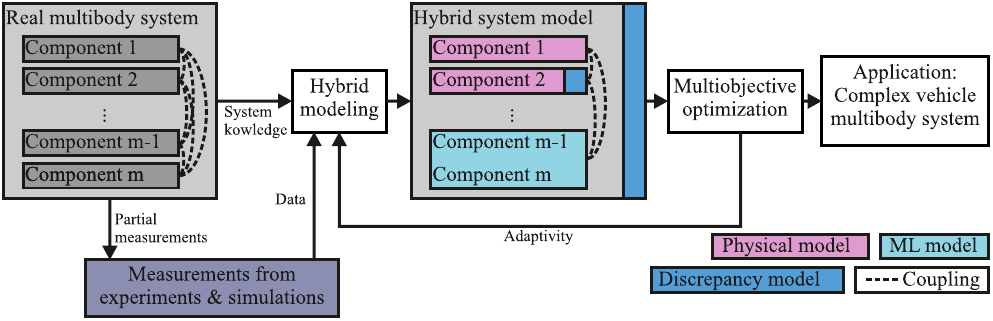}
    \caption{Approach to assist the multi-criteria design of complex multibody systems using adaptive hybrid models.}
    \label{fig:HyM3}
\end{figure}

The main goal of this project is to develop a highly flexible and adaptive data-enhanced framework for the multi-criteria design of complex multibody systems, see Fig.~\ref{fig:HyM3}. In the design of new systems, various (usually conflicting) goals and different parameters can be adjusted to meet these goals. For optimization, simulations must be carried out frequently, which can be computationally intensive and time-consuming, especially for complex multibody systems. During the design process, it is important to maintain the transparency of the system and correctly map dependencies between sub-models.
\begin{enumerate}
    \item \textbf{Prior knowledge/experience:} During the design of new dynamical systems, a typical approach involves iteratively defining requirements, creating conceptual designs, developing detailed models incorporating both physical and mathematical aspects, and refining the design through iterative analysis and simulation to meet performance goals and constraints. Finally, the design is validated through testing to ensure compliance with requirements and reliable operation.
    \item \textbf{Model:} A physical model is constructed based on first principles. Initially, a topology is defined, which is then supplemented with force elements. These elements can be very complex and nonlinear, making precise modeling and parameterization challenging, if not impossible. For the most accurate modeling, the integration of flexible bodies may also be necessary, though this comes with high computational costs. Parameters of the physical model can be identified through measurements of individual components or the entire system. Design variables are defined within the physical model and are adjusted to fulfill the defined objective function. Subsequently, multi-objective optimization is performed using the established model.
    \item \textbf{Limits and boundaries:} Challenges in modeling often arise from the need to simplify complex real-world systems, leading to potential inaccuracies. Incomplete or uncertain data may hinder accurate model creation, and errors in assumptions or equations can compromise reliability. Managing the complexity of intricate systems presents difficulties in analysis, simulation, and implementation. The absence of precise modeling approaches, unaccounted for or unknown phenomena, and the complexity of parameter identification, particularly for sub-systems that remain unchanged, are significant challenges.
    In terms of optimization, the resulting model may be prohibitively expensive to evaluate, or it may be hard to obtain gradient information.
    \item \textbf{Data:} Simulation data generated by a physical model in multibody dynamics simulation software (Adams) serve as the foundation for understanding the system's behavior. Additionally, measurements of sub-systems/components are used for parameterizing individual components, enhancing model accuracy. Measurements on the entire system are essential for validating overall system performance, ensuring that simulated behavior aligns with real-world observations.
\end{enumerate}
The development of a highly flexible and adaptive data-enhanced framework is expected to significantly improve the efficiency and accuracy of the multi-criteria design of complex multibody systems \cite{BP21}. By effectively integrating physical and data-driven models, the framework aims to reduce computational costs, maintain system transparency, and ensure accurate representation of dynamic behaviors, ultimately leading to optimized design outcomes \cite{Wohlleben2022,Wohlleben2023,Wohlleben2024}.

\subsection{Reducing noise emissions from and in mechanical structures (Project of S.C. Langer and T. Lüddecke)}\label{sec:acoustics_project}

The aim of this project is to develop tools and methods for reducing noise emitted from mechanical structures, that might impact a passenger or bystander. We work on leveraging deep learning methods developed in the area of generative modeling as well as deep learning based surrogate modeling.

\begin{enumerate}
\item \textbf{Prior knowledge/experiences:} The (vibro-)acoustic properties of mechanical structures are often not the primary purpose of a product. Because of this, acoustic issues, such as unfavorable resonances, are often addressed by comparatively inefficient and expensive additions only at the final stages of product design. Acoustic design measures also have to work within the constraints imposed by the primary purpose of a structure (e.g., an airplane must still fly).
Often the dynamic response in specific frequency bands need to be reduced, e.g., to avoid acoustic coupling with resonances of an engine.
\item \textbf{Model:} The modeling of acoustic responses in the context of stationary dynamical systems is often performed in the frequency domain. This means, not the solution for a specific time point, but rather the steady state solution assuming a constant or harmonic excitation is computed. In our project, we mainly focus on the prediction of the vibration patterns of harmonically excited plates and their acoustically-oriented design. Plates are ubiquitous elements in mechanical structures such as cars, planes or trains. They often transmit sound produced by e.g., a motor from a compartment to a passenger area. The vibrations of these plates are often simulated via the finite element method based on a geometry and physical parameters (boundary conditions, material properties, loading) of the plates. 
\item \textbf{Limits and boundaries:} 
While finite element simulations can produce accurate results, a major limitation for designing mechanical structures is the computational cost associated with such wave-resolving simulations. This is especially true at high frequencies of interest, as they require finer discretizations.
Firstly, deep learning based surrogate models offer the potential to accelerate simulations by a factor of several orders of magnitude \cite{azizzadenesheli2024neural}. Thus, candidate designs can be evaluated faster, independent of the design algorithm. Secondly, neural networks are differentiable, allowing to directly obtain gradients in the input or design space. This enables cheap and efficient design optimization and inverse design, as favorable directions in the search space are indicated by the gradients.

Additionally, data-driven generative modeling methods \cite{goodfellow14,Kingma2013AutoEncodingVB,Ho2020DenoisingDP} have shown great potential to produce samples with desired properties. These methods can be used to generate designs that fulfill acoustic constraints as well as other constraints, such as manufacturability or shape. 
In the future, a challenge will be incorporating a representation of complex 3D geometries and materials into the neural network.
\item \textbf{Data:} We published a dataset consisting of the vibroacoustic response of 12,000 harmonically excited plates \cite{delden2023vibroacoustic}. These plates have varying beading patterns. Beading patterns are indentations in plates that locally change the stiffness of the plate and can thus be used to change the vibroacoustic response. This dataset has been used for surrogate modeling and in the context of using denoising diffusion generative models for designing plates for desired properties \cite{delden2024minimizing}.
\end{enumerate}

We aim to extend our work on using advanced deep learning tools for surrogate modeling and design optimization in acoustics by investigating methods to enhance data efficiency for surrogate modeling through active learning.

\subsection{
AI-design assistant for legged robots and other periodically operating mechanisms (Project of C.D.~Remy)}\label{sec:LegDesign}

The aim of this project is to develop a systematic design optimization methodology for mechanisms and machines that perform periodic motions.
The design goal for these systems is to minimize an effort-based cost which is evaluated across various periodic motions.
The design framework is based on a numerical co-optimization of motion trajectories and mechanical design parameters.
\vspace{-4mm}
\begin{figure}[bth]
    \centering
    \includegraphics{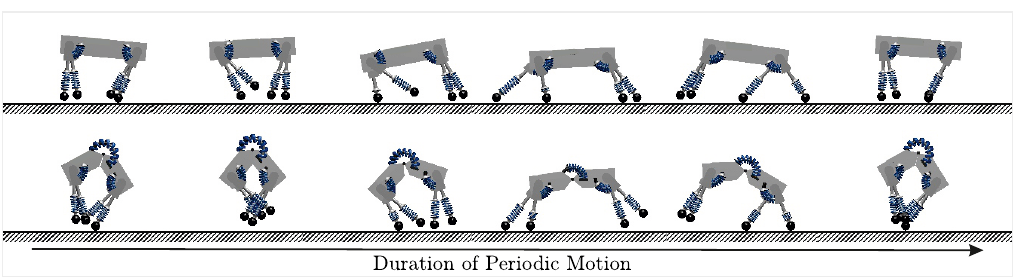}
    \caption{Illustration of how mechanical dynamics inherently couple a robotic system's design (here: rigid or flexible spine) and the resulting optimal motion (here: a galloping gait). Figure adapted from~\cite{Yesilevskiy2018}.}
    \label{fig:leggedSystem}
\end{figure}
\vspace{-2mm}
\begin{enumerate}
\item \textbf{Prior knowledge/experiences:} 
The primary challenge in such a design optimization lies in the inherent coupling between a system's design parameters and its mechanical dynamics.
Any change in parameters affects the system's dynamics, impacting the optimal motion and the overall performance cost~\cite{yesilevskiy2018energy}.
This coupling is illustrated by the key-frames in Fig.~\ref{fig:leggedSystem}, which show the periodic motions of a quadrupedal robot with two distinct spinal designs~\cite{Yesilevskiy2018}.
Additionally, designing for a wide range of periodic motions increases the complexity of an already high-dimensional, highly non-convex optimization problem, which is labor-intensive to solve in its current form.

\item \textbf{Model:} 
The mechanical models under investigation include pick-and-place mechanisms and legged robotic systems. 
These systems are modeled as hybrid dynamical systems, comprising continuous-time dynamics and discrete-time mappings that represent interactions with the environment (e.g., foot touchdown). 
These non-smooth and nonlinear dynamics are the primary source of non-convexity in the optimization problem, posing a challenge in constructing effective gradient descent directions.

\item \textbf{Limits and boundaries:} 
Addressing the inherent coupling between parameter design and dynamics involves formulating the co-design problem as a bi-level optimization challenge. In this framework, design parameters are initially fixed during lower-level trajectory optimization involving various periodic motions, and subsequently adjusted in an upper-level step.
When exploring larger task or parameter spaces, numerical continuation techniques can help reduce the computational burden~\cite{raff2022generating, rosa2023approach}.
The challenge of finding good minimizers can be further addressed by replacing the hybrid dynamics with learned surrogate models. 
The difficulty here lies in training surrogate models that accurately represent non-smooth dynamics while facilitating valuable gradient information for the design optimization process.

\item \textbf{Data:} 
The data for the surrogate models are primarily obtained from targeted sampling of the actuation- and state-space. 
To partially mitigate the curse of dimensionality in data sampling, we aim to include data from optimization iterations that result from gradient descent on the original problem.
\end{enumerate}

This project aims to develop a design assistant tool capable of swiftly identifying the optimal parameter set for any given system. 
This advancement will lead to more efficient robotic designs capable of operating across a spectrum of periodic motions.

\subsection{AI enabled design for dynamics in the small data limit (Project of M.~Stender)}\label{sec:TUBerlin}

\begin{figure}[bth]
    \centering
    \includegraphics[width=\textwidth]{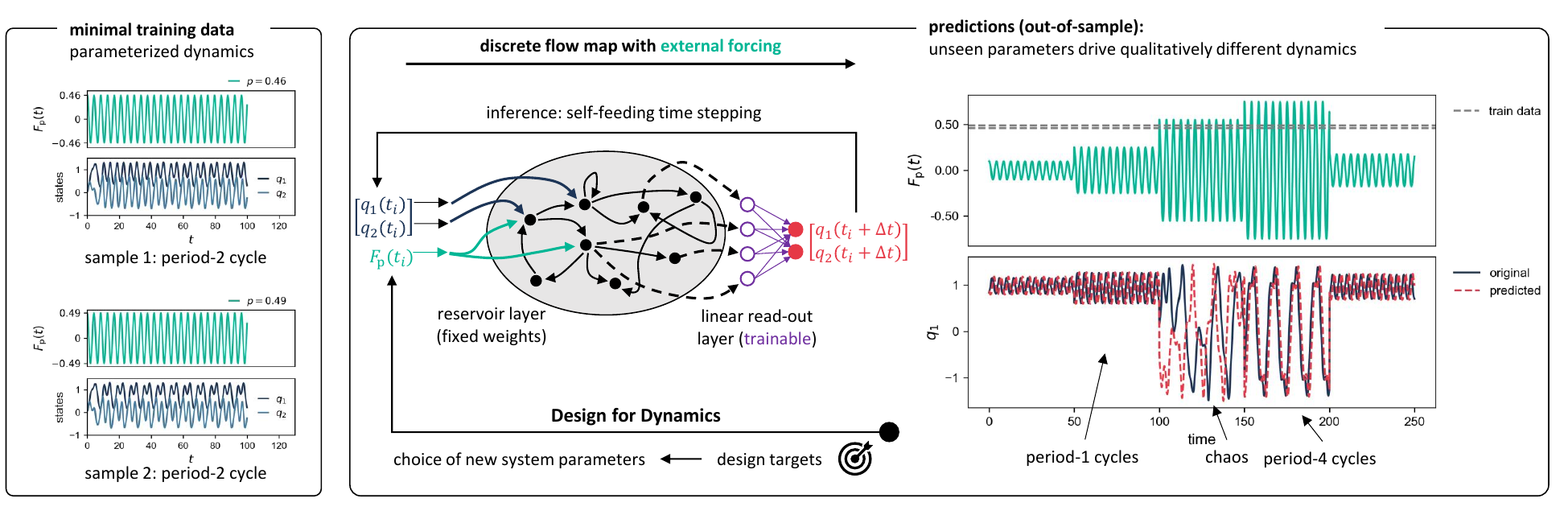}
    \caption{Reservoir computing (RC) and design for dynamics: RC allows for learning from a minimal data set (left panel) of forced nonlinear dynamics. Out-of-sample generalization behavior is obtained that generates predictions of qualitatively unseen dynamics under varying system parameters $p$.}
    \label{fig:TUB_case}
\end{figure}

Mechanical engineering design often simplifies system dynamics, treating non-stationary external loads as minor deviations from steady-state. However, real-world systems like aircraft engines and wind turbines experience complex, non-stationary, multi-physical loads, leading to equally complex responses. Accounting for nonlinear dynamics and critical transitions due to system parameter variations is crucial in the early design process. This project leverages a resource- and data-efficient Reservoir Computing framework to predict parameterized dynamics, using a Duffing-type nonlinear oscillator as an example (see Figure~\ref{fig:TUB_case}). 

\begin{enumerate}
    \item \textbf{Prior knowledge/experience:} Prior knowledge in technical design can vary widely. In some cases, well-defined loads and conditions guide the design process, such as avoiding resonance. In others, unpredictable dynamic loads necessitate evaluating a broad range of conditions. For the given project, the nonlinear effects dominate the system response. A challenge arises from qualitative changes in the system dynamics (bifurcations) triggered by minor adjustments to system parameters. As a result, prior knowledge from similar designs may become irrelevant due to new dynamic behaviors induced by small variations in parameters \cite{thompson2002nonlinear}.
    \item \textbf{Model:} The mechanical oscillator studied is governed by a parametrized ordinary differential equation in terms of external forcing amplitude and frequency, exhibiting rich dynamics across different parameter values. While semi-analytical solutions are possible, the model represents a broader class of nonlinear dynamical systems requiring optimal parameter selection in design for required dynamics. Typically, analysis relies on linearization methods like modal analysis \cite{ewins2009modal}. Direct time integration can simulate nonlinear dynamics but is computationally expensive and limited by model quality and resource demands. Artificial neural networks are widely used for predicting system dynamics under varying parameters \cite{stender2021deep, stender2023u}, but their high resource demands are a limitation. This project instead utilizes Reservoir Computing \cite{jaeger2002adaptive}, a recurrent network model. It employs a stateful, unstructured, non-trainable reservoir for input representation, requiring only a linear output layer trained via regression. The stateful reservoir is ideal for capturing system dynamics, as these are inherently integrated by design.  
    \item \textbf{Limits and boundaries:} Linearized models fail when nonlinearities arise (e.g., contact interfaces, nonlinear material behavior), and simulation models are constrained by simplifying assumptions. Given the incomplete understanding of processes like frictional contacts and multi-physical interactions, simulation discrepancies are expected. Data-driven models face different limitations: as general function approximators, they often lack adherence to physical principles such as energy conservation. Their generalization depends heavily on the quality and quantity of training data. The proposed Reservoir Computing framework addresses some of those limitations through its inherently dynamic learning approach. A preliminary study for the given problem demonstrates that a reservoir computer can generalize to unseen dynamics in far-out-of-distribution cases \cite{yadav2024predicting}, making it promising for predicting complex dynamics from limited observations. Current limitations to forcing parameters will be adressed in future, i.e. to include structural parameters for helping the design. 
    \item \textbf{Data:} Data in engineering is typically sparse due to the high costs of data acquisition and limited access to systems during operation. The expense and difficulty of generating large simulation datasets make data-hungry learning methods impractical. Experimental data from prototypes is also scarce and cannot be generated for every design variation. To address this, we have provided a minimal dataset\footnote{\href{https://github.com/maneesh51/Benchmark-Tasks/tree/main}{DORA open benchmark data set}} for a nonlinear mechanical oscillator, showcasing significant variations in system responses with small parameter changes. Collected via direct time integration, this dataset is ideal for benchmarking the predictive performance and data requirements of data-driven models.
\end{enumerate}

The project aims to develop (a) predictive data-driven models with high generalization capabilities for parameterized dynamics, (b) using minimal data and computational resources, and (c) assisting in the intentional optimization of system dynamics, as illustrated in Fig.~\ref{fig:TUB_case}. The efficient Reservoir Computing framework is ideal for rapid parametric design studies and load scenarios, significantly reducing time spent on simulations and experimental testing.

\section{Conclusion}
\label{sec:conclusion}
In this paper, we discuss the potential of AI-based methods in the design of technological systems and we explore available methods and application domains. We suggest to aim for AI-assisted development by intertwining model-based simulation and optimization techniques with machine learning methods throughout the V-model.
In fact, the ability of modern AI, in particular, machine learning methods, to handle large amounts of data, makes AI the ideal third pillar besides experiment-based design and simulation-based design.
Among others, data can be enriched by synthetic data provided by GANs and data-based models provide attractive alternatives as surrogates or in a hybrid modeling fashion.
Available optimization methods range from gradient-based classical approaches and genetic algorithms to reinforcement learning.
For design optimization, parameter identification and also (optimal) control, we have presented various application examples from the SPP~2353. These range from the specific design of continuum robots, crash structures and legged robots to the general design for emission reduction, technical  and multibody systems. Both the examples and the AI methods used cover a broad field and can be easily extended to other applications.  

This article can serve with best practice examples as a guide for potential, inexperienced as well as experienced users of AI in the industrial environment and provide contacts to experts.


\section*{Authors' contributions}

K.\ Flaßkamp and K.\ de Payrebrune authored Section~\ref{sec:introduction}, Section~\ref{sec:conclusion} and equally steered and organized the overall paper design.\\
T.\ Ströhla and T.~Sattel authored Section~\ref{sec:currentstate}.\\
M.\ Stoffel, R. Gulakala and A. Borse authored Section~\ref{sect:synthetic data},  Section~\ref{Reinforcement Learning} and Section~\ref{sec:vehiclecrashopt}\\
S.\ Peitz  authored Section~\ref{subsubsec:SurrogateModeling} and the fluid dynamics part within Section~\ref{sec:application_domains}\\
M.\ Wohlleben and W.\ Sextro authored Section~\ref{sec:HybridModeling} and Section~\ref{sec:HyM3}.\\
D.\ Bestle authored Section~\ref{identification, optimization and control}.\\
K.\ de Payrebrune authored Section~\ref{sec:SoftRobotics}.\\
K.\ Flaßkamp and T.\ Sattel authored Section~\ref{sec:Neurosurgery}.\\
D.\ Bestle, P.\ Eberhard and B.\ Röder authored Section~\ref{sec:BestleEberhard} and the mechanical engineering part within Section~\ref{sec:application_domains}.\\
J.\ van Delden, J. Schultz, T. Lüddecke, C. Blech and Sabine C. Langer authored the acoustics part within Section~\ref{sec:application_domains} and Section~\ref{sec:acoustics_project} .\\
M.\ Raff and C.D.\ Remy authored the robotics and mechatronics part within Section~\ref{sec:application_domains} and Section~\ref{sec:LegDesign}.\\
M.\ Stender and M. Yadav authored Section~\ref{sec:TUBerlin}.

\section*{Acknowledgements} 
This work was supported by the Deutsche Forschungsgemeinschaft (DFG, German Research Foundation) under Grant 460725022 (Priority Programme SPP 2353 ‘Daring More Intelligence – Design Assistants in Mechanics and Dynamics’). The following projects were involved: 501840485 (D.~Bestle, P.~Eberhard), 501862165 (D.~Remy), 501927736 (S.~Langer, T.~Lüddecke), 501847579 (M.~Stender), 501834605 (S.~Peitz, W.~Sextro), 501928699 (K.~Flaßkamp, T.~Sattel), 501861263 (K.M.~de~Payrebrune), 501877598 (M.~Stoffel). This support is highly appreciated.

\FloatBarrier
\bibliographystyle{naturemag}

\bibliography{references}

\end{document}